\documentclass[style=bcnpostgrad,entry]{proceedings}
\renewcommand{\conferencetitle}{Barcelona Postgrad Encounters on Fundamental Physics}
\usepackage[utf8]{inputenc}
\usepackage{graphicx}
\usepackage{caption}
\usepackage{subcaption}

\newcommand{\bea}{\begin{eqnarray}}
\newcommand{\eea}{\end{eqnarray}}

\begin{document}

\author[1,a]{\underline{Danielle Wills}}
\email[1]{d.e.wills@durham.ac.uk}
\affiliation[a]{Centre for Particle Theory, Durham University,\\ South Road, DH1 3LE Durham, United Kingdom}
\author[2,c]{Konstantinos Dimopoulos}
\email[2]{k.dimopoulos1@lancaster.ac.uk}
\affiliation[c]{Consortium for Fundamental Physics, Physics Department, Lancaster University,\\ LA1 4YB Lancaster, United Kingdom}
\author[3,b]{Ivonne~Zavala}
\email[3]{e.i.zavala@rug.nl}
\affiliation[b]{Centre for Theoretical Physics, University of Groningen,\\ Nijenborgh 4, 9747 AG Groningen, The Netherlands}

\title{D-branes and cosmic structure} 

\abstract{We outline the embedding of the vector curvaton scenario as a promising mechanism to generate statistical anisotropy within Type IIB string theory, where the vector field on a single D-brane plays the role of the vector curvaton. We first consider a toy model in the context of open string inflation, and then begin to construct a concrete model in the context of closed string inflation.} 

\acknowledgements{DW would like to thank Tomi Koivisto for insightful discussions. DW is supported by an STFC studentship. K.D. is supported by the Lancaster-Manchester-Sheffield Consortium for Fundamental Physics under STFC grant ST/J000418/1.} 

\maketitle

\section{Introduction}

Type IIB string theory has provided a rich arena for embedding phenomenological scenarios of the early universe within a fundamental theory, providing a wealth of light fields with masses and couplings that can in principle be known explicitly, which allows for concrete model-building. In such a context in which the early universe is expected to be diversely populated, it is natural to wonder whether or not the features captured in CMB maps are in fact the result of an intricate piece of team work, rather than a single lone wolf inflaton. In particular, the primordial seeds of cosmic structure are believed to be generated by gravitational particle production, a quantum process that requires an expanding background, but does not require the quantum and classical sectors to be one and the same. Pursuance along these lines of study led to the curvaton scenario \cite{Lyth:2001nq} and, more recently, vector curvaton scenario \cite{Dimopoulos:2006ms,Dimopoulos:2009vu}, in which the central idea is that the fields which give the dominant contribution to the seeds of cosmic structure have nothing to do with expanding the background (for vector field driven cosmologies, see \emph{e.g.} \cite{Koivisto:2008xf}).

Vector fields arise on the world-volumes of D-branes as a result of the open strings which end of them. Therefore, although their cosmological implications have been largely unexplored, such fields are completely generic in open string as well as closed string inflation models because all of these models contain D-branes, either at an active or passive level. In open string inflation, D-branes (most famously D3-branes) can actively drive cosmic expansion as they move in warped throat environments (see \cite{Kallosh:2007ig} for a review). In closed string inflation, K\"ahler moduli which parameterise the volumes of compact 4-cycles may drive inflation as they roll to their minima \cite{Conlon:2005jm}, and these 4-cycles may be wrapped by D7-branes, which are required to be present by consistency of the theory. In these proceedings we will explore the string constructions under which D-brane vector fields may become curvatons in both open and closed string inflation scenarios, where the former will provide a toy model and the latter will be the starting point for a concrete model.

\begin{figure}
        \centering
            \begin{subfigure}[b]{0.5\textwidth}
                \includegraphics[width=\textwidth]{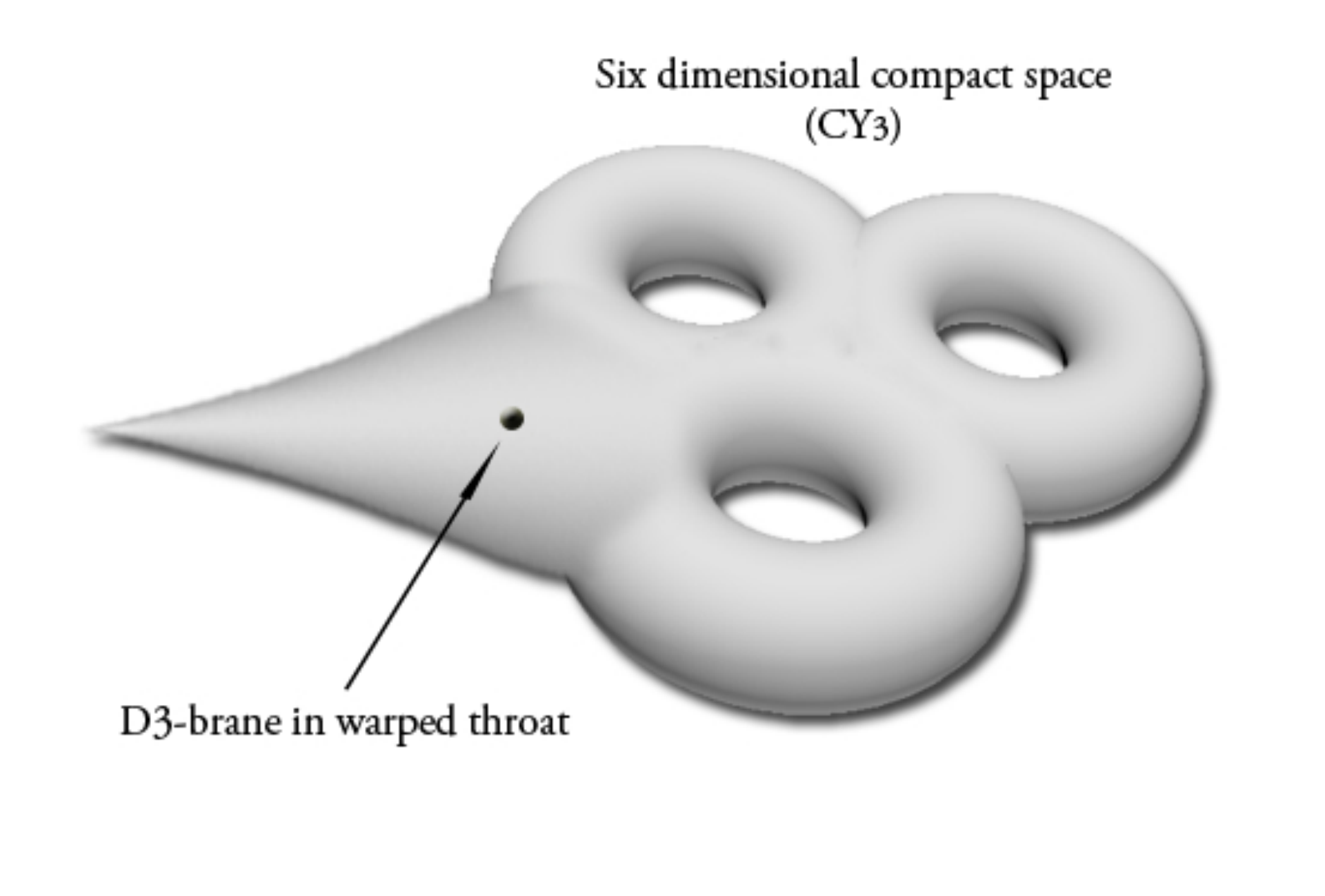}
                \label{largevolume}
            \end{subfigure}
             \begin{subfigure}[b]{0.45\textwidth}
                \includegraphics[width=\textwidth]{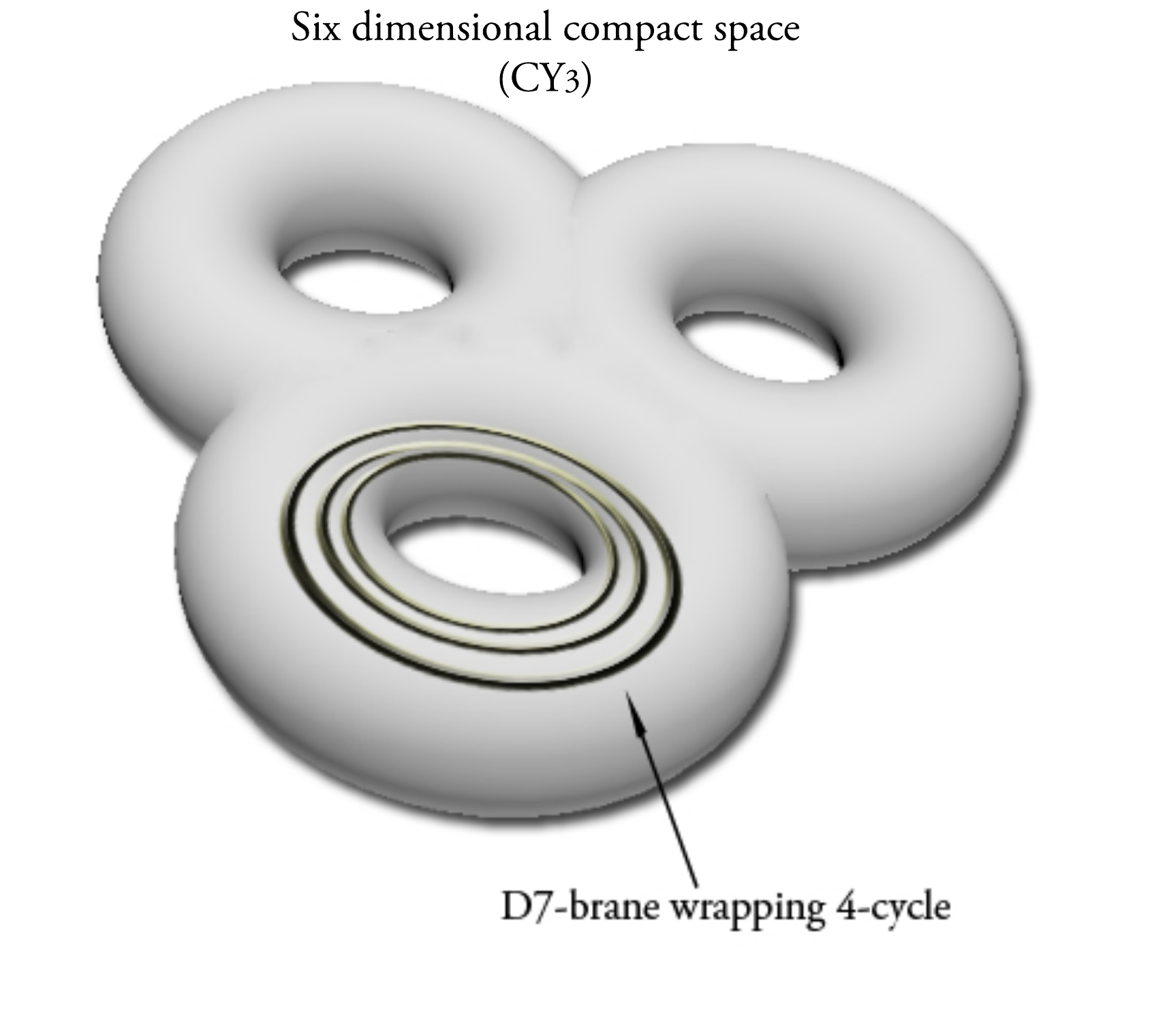}
                       \label{largevolume}
            \end{subfigure}
\end{figure}

\section{The vector curvaton scenario}\label{veccurvscen}

The seeds of cosmic structure induce tiny regions of over- and under-densities in the otherwise uniform energy density of the early universe, and at these loci the geometry is correspondingly perturbed. This gives rise to a quantity known as the curvature perturbation $\zeta$, which forms the initial condition for structure growth. The vector curvaton scenario explores the conditions under which a vector field may produce a sizable contribution to $\zeta$.

In order to undergo gravitational particle production, fields must be sufficiently light such that their Compton wavelengths may extend beyond the horizon (plus not be conformally invariant). In order to be viable curvatons, they must also give rise to spectra that are consistent with observations of $\zeta$ and thus, importantly, suitably scale-invariant. In one version of the vector curvaton scenario \cite{Dimopoulos:2009am}, it is shown that a light vector field with a mass and a gauge kinetic function that varies during inflation may give rise to a scale-invariant spectrum of superhorizon perturbations if the mass and gauge kinetic function obey
\begin{equation}\label{vectorcons1}
m \propto a(t),\hspace{1cm}f\propto a(t)^{2}
\end{equation}
or
\begin{equation}\label{vectorcons2}
m \propto a(t),\hspace{1cm}f\propto a(t)^{-4},
\end{equation}
where $a(t)$ is the cosmic scale factor. To see this\footnote{We summarise here the computations done in \cite{Dimopoulos:2009am}.}, consider the Lagrangian for a massive Abelian vector field,
\begin{equation}
\mathcal{L}= -\frac{1}{4}f F_{\mu\nu}F^{\mu\nu} - \frac{1}{2}m^2 A_\mu A^\nu,
\end{equation}
where $F_{\mu\nu}= \partial_\mu A_\nu - \partial_\nu A_\mu$ is the field strength tensor, and $f$ and $m$ are the gauge kinetic function and mass respectively. We assume that $f\propto a(t)^\alpha$ and $m\propto a(t)^\beta$, since $a(t)$ sets the only time scale in the problem. For  quasi- deSitter expansion with Hubble parameter $H$, we find the following equations of motion for the homogeneous $A_\mu(t)$ where $A_\mu  = (A_0, \boldsymbol{A})$:
\begin{equation}\label{vectorEOM1}
A_0 = 0,
\end{equation}
\begin{equation}\label{vectorEOM2}
\boldsymbol{\dot{A}}+ \left(H+\frac{\dot{f}}{f}\right)\boldsymbol{\dot{A}}+\frac{m^2}{f}\boldsymbol{A}=0.
\end{equation}
It is clear that the effective mass of the vector field is given by
\begin{equation}
M \equiv \frac{m}{\sqrt{f}},
\end{equation}
where $M \ll H$ while the cosmological scales exit the horizon. To study gravitational particle production we perturb around the homogeneous zero-mode $A_{\mu}(t)$,
\begin{equation}
A_{\mu}(t,\boldsymbol{x}) = A_{\mu}(t)+ \delta A_{\mu}(t,\boldsymbol{x}),
\end{equation}
and we define the canonically normalised physical (as opposed to comoving) vector field as
\begin{equation}
W = \sqrt{f}A/a
\end{equation}
with perturbations $\delta W$. In momentum space, the equations of motion for the left and right transverse and longitudinal polarisations of the quantum mode functions of $\delta W$, $w(t,\boldsymbol{x})$, are found to be
\begin{equation}
\ddot{w}_{L,R}+3H\dot{w}_{L,R}+\left[-\frac{1}{4}(\alpha +4)(\alpha -2)H^2  + \left(\frac{k}{a}\right)^2\right]w_{L,R}=0,
\end{equation}
\begin{equation}
\ddot{w}_{||}+(5-\alpha+2\beta)H\dot{w}_{||}+ \left[-\frac{1}{2}(\alpha-2)(2 - \alpha+ 2\beta)H^2+\left(\frac{k}{a}\right)^2\right]w_{||}=0,
\end{equation}
where we have assumed $M \ll H$. We may now compute the superhorizon power spectra for these polarisations. For the left and right transverse modes, we find
\begin{equation}\label{powerLR}
\mathcal{P}_{L,R}= \frac{k^3}{2\pi^2}\left|\lim_{k/aH \rightarrow 0^+}w_{L,R}\right|^2 \propto \left(\frac{H}{2\pi}\right)^2\left(\frac{k}{2aH}\right)^{3-2\sigma}
\end{equation}
where $\sigma = \frac{1}{2}|\alpha + 1|$.
For the longitunal mode,
\begin{equation}\label{powerLong}
\mathcal{P}_{||}= \frac{k^3}{2\pi^2}\left|\lim_{k/aH \rightarrow 0^+}w_{||}\right|^2 \propto \left(\frac{3H}{M}\right)^2 \left(\frac{H}{2\pi}\right)^2\left(\frac{k}{2aH}\right)^{5-2\rho}
\end{equation}
where $\rho=\frac{1}{2}\sqrt{9+2(\alpha+1)(2-\alpha+2\beta)+(2-\alpha+2\beta)^2}$ and $M$ enters via a Lorentz boost.

We then see from Eq.~(\ref{powerLR}) that the transverse spectrum is scale-invariant if $\alpha = -(1\pm 3)$. From Eq.~(\ref{powerLong}), we see that the longitudinal spectrum is scale-invariant if, in addition, $\beta = -1/2(3\pm 5)$, but we reject $\beta =-4$ because $M$ decreases with $a$ for this case and so $M \ll H$ cannot hold in the subhorizon limit. Hence we arrive at Eqs.~(\ref{vectorcons1},\ref{vectorcons2}). Note that a necessary condition to obtain these relations is that $f$ and $m$ have the precise relationships
\begin{equation}\label{a2case}
f \propto  m^{2}
\end{equation}
or
\begin{equation}\label{a4case}
f\propto m^{-4},
\end{equation}
respectively, in order for the vector field to generate scale-invariant spectra for all of its components.

\section{D-brane vector curvatons}

The excitations of D-branes are described by the oscillation modes of the open strings that end on them. The massless bosonic open string spectrum includes a U(1) gauge boson which propagates only along the world-volume, and there are massless scalars which describe the fluctuations in the position of the brane in the transverse directions. The U(1) world-volume vector fields may then obtain masses via St\"uckelberg couplings to bulk two-forms, and these masses can depend on the various scalars in the theory. In addition, the vector fields may couple to the scalars through their gauge kinetic functions. If the scalars that enter these quantities are evolving at inflationary energies, then the early universe would contain massive D-brane vector fields with time varying masses and gauge kinetic functions. Then, if the vectors are light and the dynamics of the scalars obey Eqs.~(\ref{vectorcons1},\ref{vectorcons2}) while the cosmological scales exit the horizon, these D-brane vector fields could become vector curvatons and contribute to or even generate the seeds of cosmic structure. We will now examine these possibilities.

\subsection{The general set-up}

We consider a warped flux compactification of Type IIB string theory in which the ten dimensional metric takes the form
\begin{equation}
G_{MN}dx^Mdx^N=h^{-1/2}g_{\mu\nu}dx^\mu dx^\nu +h^{1/2}g_{AB}dy^A dy^B,
\end{equation}
where the indices $M,N = (0,...,9)$, $\mu,\nu = (0,...,3)$ and $A,B=(4,...,9)$ denote coordinates in ten dimensional spacetime, the four noncompact dimensions and the six compact dimensions respectively, and $h$ is the warp factor which depends only on the compact coordinates. We embed a probe D$p$-brane with world-volume coordinates $\xi^a$ in this background, with three of its directions extended in the three large spatial dimensions, and its remaining $(p-3)$ directions wrapping a compact $(p-3)$-cycle. The action for such a brane is a sum of the Dirac-Born-Infeld (DBI) and Wess-Zumino (WZ) actions, where the former encodes kinetic terms for the brane and its world-volume fields, and the latter, couplings of world-volume fields and the brane itself to other fields in the bulk. In the Einstein frame, the DBI action takes the form
\begin{equation}\label{genDBI}
S_{DBI}= -\mu_{p}\int d^{p+1}\xi e^{\frac{(p-3)}{4}\phi}\sqrt{-\det(\gamma_{ab}+e^{-\frac{\phi}{2}}\mathcal{F}_{ab})}.
\end{equation}
In this expression, $\phi$ is the dilaton which parameterises the string coupling, and the D$p$-brane tension $T_p=\mu_p e^{((p-3)/4)\phi}$ where $\mu_p=(2\pi)^{-p}(\alpha')^{-(p+1)/2}$ and $\alpha'=\ell_s^2$ is the string scale. In addition,
\begin{equation}
\mathcal{F}_{ab}=\mathcal{B}_{ab}+ 2\pi \alpha' F_{ab},\hspace{1cm}\gamma_{ab}=G_{MN}\partial_a x^{M}\partial_b x^{N}
\end{equation}
where $\mathcal{B}_{ab}$ is the pullback of the NSNS 2-form field and $F_{ab}$ is the field strength of the world volume gauge field, and $\gamma_{ab}$ is the pullback of the ten dimensional metric on the brane.

The WZ action takes the form
\begin{equation}\label{genWZ}
S_{WZ}=q \mu_p \int_{\mathcal{W}_{p+1}}\sum_n \mathcal{C}_n \wedge e^\mathcal{F},
\end{equation}
where the $\mathcal{C}_n$ are the pullbacks of bulk fields $C_n$ on the brane and $q=+1(-1)$ for a probe D$p$-brane(anti-brane). The integral is over the world-volume $\mathcal{W}_{p+1}$ and the sum is over the relevant rank $(p+1)$ products of bulk and world-volume fields.

We work in the static gauge in which $\xi^\mu = x^\mu$. For the compact spacetime coordinates transverse to the brane (with indices $i,j = (p+1,...,9)$), we allow $y^i=y^i(\xi^\mu)$: these functions will give the massless open string modes that parameterise the fluctuations in the position of the brane.

\subsection{D3-brane vector curvaton}

We will first consider the vector field on a D3-brane in the context of open string inflation \cite{Dimopoulos:2011pe}. In these inflation models, the inflaton is identified with the changing position coordinate of a D$p$-brane as it moves along its potential in a warped throat. If the potential is sufficiently flat, the brane moves slowly and the scalar position field which parameterises the motion has a nearly constant energy density, giving rise to an epoch of slow-roll inflation. If the potential is steep on the other hand but the brane is moving in a strongly warped region, a surprising result is that the energy density still continues to be nearly constant. This is due to the fact that the strong warping forces the brane to slow down despite the steepness of the potential, and once again this gives rise to an epoch of inflation, in this case known as DBI inflation \cite{Alishahiha:2004eh}. The majority of models consider D3-branes however other possibilities have been studied.

From the form of Eq.~(\ref{genDBI}) we see immediately that the gauge kinetic function for $A_\mu$ contains a dynamical degree of freedom, namely the dilaton $\phi$. Then, for a D3-brane, the WZ action in Eq.~(\ref{genWZ}) contains a coupling $C_2 \wedge F_2$ which is able to generate a mass for the vector field via the St\"uckelberg mechanism\footnote{For compactifications of Type IIB with O3/O7 orientifold planes, the four dimensional components of $C_2$ are projected out of the spectrum, however we will continue to use this coupling for the purpose of the toy model.}. Computing the mass explicitly yields
\begin{equation}\label{3mass}
m = e^{-\phi/2}\sqrt{\pi}(2\pi)^5 \frac{M_P}{\mathcal{V}_6},
\end{equation}
where $\mathcal{V}_6$ is the dimensionless volume of the compact space and $M_P$ is the Planck mass. Therefore, both the mass and gauge kinetic function for the vector field contain the dilaton. While this field is usually considered to be stabilised during inflation, for the purpose of our toy model we will assume that it can be dynamical, at least while the cosmological scales exit the horizon.

The total four dimensional action for the D3-brane that we consider, including all appropriate terms for gravity, the evolving dilaton $\phi$, the canonically normalised inflaton $\varphi$ and the canonically normalised vector curvaton $A_\mu$ (after expanding the determinant in Eq.~(\ref{genDBI})\footnote{We keep terms only up to quadratic order in $F$ and $\partial_\mu \phi$ and in their products for now, see \cite{Dimopoulos:2011pe} for the more general case.} and taking into account the mass generation mechanism from Eq.~(\ref{genWZ})), then takes the form
\begin{align}\label{totaction3}
\begin{split}
S_{D3}= &\int d^4 x \sqrt{-g}h^{-1}\left(\frac{M_P^2}{2}R - \frac{M_P^2}{4}\partial_\mu\phi \partial^\mu\phi - V(\phi)\right.\\
&\left.-\left[1+\frac{1}{2}he^{-\phi}F_{\mu\nu}F^{\mu\nu}+h\partial_\mu \varphi \partial^\mu \varphi\right]^{1/2}-V(\varphi)+h^{-1}-\frac{1}{2}m^2 A_\mu A^\mu\right).
\end{split}
\end{align}
We consider the fields to evolve in an FRW universe, in which case the equation of motion for $\varphi(t)$ is given by
\begin{equation}\label{inflatonEOM}
\ddot\varphi - \frac{h'}{h^2}+ \frac{3}{2}\frac{h'}{h}\dot\varphi^2 + 3H\dot\varphi\frac{1}{\gamma_\varphi^2}- \frac{h'}{h}e^{-\phi}\left(\frac{\dot A}{\gamma_\varphi a}\right)^2 + \left(V'(\varphi)+ \frac{h'}{h^2}\right)\frac{1}{\gamma_\varphi^3}=0,
\end{equation}
where
\begin{equation}
\gamma_\varphi = \frac{1}{\sqrt{1-h\dot\varphi^2}}
\end{equation}
is the Lorentz factor for the inflaton. For a Maxwellian vector field we may expand the square root in Eq.~(\ref{totaction3}) such that the equations of motion for $A_\mu(t)$ take the form given in Eqs.~(\ref{vectorEOM1},\ref{vectorEOM2}), with the mass $m$ given by Eq.~(\ref{3mass}), and the gauge kinetic function $f$ still to be determined.

We first focus on the simplest possibility: the vector field on a stationary D3-brane, in which case inflation is driven by, for example, the motion of another D3-brane. The gauge kinetic function is then
\begin{equation}\label{gauge1}
f=e^{-\phi},
\end{equation}
and the vector field decouples from the dynamics of the inflaton. Taking into account the precise powers of the dilaton $\phi$ that appear in Eqs.~(\ref{3mass}) and (\ref{gauge1}), we see that Eq.~(\ref{a2case}) is obtained. For our toy model we will merely assume that, in addition, Eq.~(\ref{vectorcons1}) may hold\footnote{This behaviour requires a linear potential for the dilaton, which could be an approximation to an exponential potential for small displacements. However, this behaviour should hold for at least the 10 efolds that span the cosmological scales, which is unfortunately not a small displacement.}, and move on to consider the cosmology. It is interesting though that the very precise relationship between $f$ and $m$ in Eq.~(\ref{a2case}) does in fact happen to arise for the stationary D3-brane vector field. The cosmological features of the vector curvaton on a stationary D3-brane have been computed in \cite{Dimopoulos:2011pe}, where it is shown that statistical anisotropy in both the spectrum and the bispectrum of the curvature perturbation can arise from such a scenario. The amount of statistical anisotropy in the spectrum, parameterised by $g$, is found to be
\begin{equation}
\sqrt{g} \sim \frac{\Omega_A}{\zeta}\frac{\delta W}{W}\sim 0.1,
\end{equation}
where $\Omega_A$ is the density parameter for the vector field\footnote{Observations suggest $g\leq 029$ \cite{Groeneboom:2009cb}. Statistical anisotropy will be observed by the Planck satellite if $g\geq 0.02$ \cite{Pullen:2007tu}.}. For the bispectrum, the non-linearity parameter $f_{\rm NL}$ is found to be
\begin{equation}
12 \ \leq \ \parallel f_{\rm NL}^{\rm eq} \parallel_{\rm max} \ \leq \ 713,
\end{equation}
while the degree of statistical anisotropy in the bispectrum, parameterised by $\mathcal{G}$, is
\begin{equation}
\mathcal{G}=\frac{1}{8}\left(\frac{3H_\star}{M}\right)^4 \gg 1,
\end{equation}
which shows that non-Gaussianity is predominantly anisotropic.

We now consider the vector field on a moving D3-brane in a warped throat. For simplicity we consider the throat to be AdS-like in the regions of interest, in which case the warp factor takes the simple form $h = \lambda/\varphi^4$ where $\lambda$ is the t'Hooft coupling. In this case we find for the gauge kinetic function
\begin{equation}
f = e^{-\phi}\gamma_\varphi.
\end{equation}
Now, for slow-roll inflation, $\gamma_\varphi \rightarrow 1$ and the results obtained for the stationary D3-brane case follow here. For the case of DBI inflation on the other hand, it is shown in \cite{Dimopoulos:2011pe} that
\begin{equation}
f \rightarrow a(t)^{2(1+\epsilon)}
\end{equation}
where $\epsilon$ is the generalised slow-roll parameter for DBI inflation. This adds a small degree of scale dependence to the spectrum, but once again the results for the stationary D3-brane follow in this case.

\subsection{D7-brane vector curvaton}

We will now investigate the vector fields on D7-branes in closed string inflation, and thereby construct a starting point for a concrete model of the vector curvaton scenario in string theory. We consider the effective action for $N=1$, $d=4$ supergravity, which takes the form
\begin{equation}\label{sugra}
S =  \int d^4x \sqrt{-g}\left(\frac{1}{2}R -\mathcal{K}_{i\bar{j}}\partial_\mu T^i \partial^\mu \bar{T}^{\bar{j}}- V(T^l,\bar{T}^{\bar{l}})\right),
\end{equation}
where the $T_i = \tau_i + i\theta_i$ are complex K\"ahler chiral fields consisting of the 4-cycle volume moduli $\tau_i$ and their associated axions $\theta_i$, and $\mathcal{K}_{i\bar{j}}$ is the K\"ahler metric. In the Large volume scenario (LVS) \cite{Conlon:2005ki}, perturbative corrections are added to the K\"ahler potential and non-perturbative corrections are added to the tree level superpotential, which stabilise the K\"ahler moduli. The potential exhibits exponentially flat directions which supports slow-roll inflation as the moduli roll to their minima \cite{Conlon:2005jm}. For a single volume modulus $\tau_m$ for example, the potential takes the form
\begin{equation}\label{kahlerpot}
V = \frac{3 W_0^2 \xi}{4\mathcal{V}^3} - \frac{4 W_0 a_m A_m \tau_m e^{-a_m\tau_m}}{\mathcal{V}^2}.
\end{equation}
Here $\mathcal{V}$ is the volume of the six dimensional Calabi-Yau in units of the string length $\ell_s$, $W_0$ is the tree level superpotential, $\xi$ is proportional to the Euler characteristic of the manifold, the $A_m$ encode threshold corrections, and the $a_m$ are constants which depend on the specific non-perturbative mechanism ($a_m = 2\pi/g_s$ for Euclidean D3-brane instantons and $a_m = 2\pi/g_sN$ for gaugino condensation on D7-branes). For a single evolving field, the canonically normalised inflaton field $\chi_m$ is related to $\tau_m$ by
\begin{equation}\label{canonicaltau}
\chi_m = \sqrt{\frac{4\lambda}{3\mathcal{V}}}\tau_m^{3/4}.
\end{equation}
The volume $\mathcal{V}$ and the 4-cycle moduli $\tau_i$ may be expressed in terms of the 2-cycle moduli $t_i$ according to
\begin{equation} \label{4cycles}
\mathcal{V}= \frac{1}{6}\kappa_{ijk}t^i t^j t^k, \quad \tau_i = \frac{\partial}{\partial t^i}\mathcal{V}= \frac{1}{2}\kappa_{ijk}t^j t^k,
\end{equation}
respectively, where $\kappa_{ijk}$ are the triple intersection numbers and $i=1,...,n=h^{1,1}$.

Geometries featuring an over-all size $\mathcal{V}$ controlled by one large 4-cycle with modulus $\tau_1$, and then a number of small blow-up 4-cycles or ``holes'' with moduli $\tau_2,...,\tau_n$, are referred to as ``Swiss-cheese'' (SC) geometries. In an appropriate basis, one can write $\mathcal{V}$ in a particularly simple diagonal form in terms of the 4-cycle moduli\footnote{In a study carried out in \cite{Gray:2012jy} which scanned hundreds of geometries, it is shown that a large number of SC geometries are expressible in the ``strong cheese'' form in Eq.~(\ref{strongcheese}).},
\begin{equation}\label{strongcheese}
\mathcal{V} = \tilde\alpha\left(\tau_1^{3/2} - \sum_{i=2}^n \lambda_i \tau_i^{3/2}\right)
\end{equation}
where $\tilde\alpha$ and $\lambda_i$ are model-dependent constants. In the LVS, geometries are chosen such that $\tau_1 \gg \tau_{2,...,n}$, therefore $\mathcal{V}$ is not destabilised if one or more of the small 4-cycles evolves.

We consider a D7-brane wrapping a 4-cycle with modulus $\tau$, which is evolving during inflation and may be driving inflation \cite{Wills:2011tk}. As for the D3-brane case, the action for such a brane descends from the general DBI and WZ actions in Eqs.~(\ref{genDBI}) and (\ref{genWZ}), where now we may integrate the DBI action over the internal components of the brane which wrap the compact 4-cycle. After expanding the determinant in the DBI action, this integration amounts to adding the 4-cycle modulus into the gauge kinetic function of the vector field. The WZ action for the D7-brane contains a coupling of the form
\begin{equation}\label{7WZ}
\frac{1}{2}(2\pi\alpha')^2\mathcal{C}_4\wedge F_2 \wedge F_2,
\end{equation}
where $C_4$ is expanded in the cohomology basis of the manifold. In this expansion is a four-dimensional 2-form $D_2$ which will give rise to a coupling of the form $D_2\wedge F_2$ when one of the field strengths in Eq.~(\ref{7WZ}) is a compact flux. This coupling generates a St\"uckelberg mass for the D7-brane vector field\footnote{Unlike $C_2$ (and indeed $B_2$) in the D3-brane case, $D_2$ is not projected out of the spectrum by the action of the orientifold planes for compactifications with O3/O7 planes.}. Again one must integrate over the compact components wrapped by the brane, and this introduces the 4-cycle modulus into the mass for the vector field. In a suitable basis, the mass is given by\footnote{The form of the the denominator arises from the integration over the harmonic 2-forms which yields
\begin{equation}
k_{ijk}t^k - \frac{(1/2k_{ilm}t^{l}t^{m})(1/2k_{ino}t^{n}t^{o})}{\mathcal{V}},
\end{equation}
and then moving to an appropriate basis as is used in Eq.~(\ref{strongcheese}).}
\begin{equation}\label{7mass}
m \propto \frac{M_P/\sqrt{\mathcal{V}}}{\sqrt{\tau^{1/2}-\frac{\tau^2}{\mathcal{V}}}}.
\end{equation}
So we see that both $f$ and $m$ for the D7-brane vector field contain the evolving 4-cycle modulus $\tau$. We may now treat the dilaton as fixed and use instead the modulus $\tau$ as the varying degree of freedom relevant for the vector curvaton scenario. As such, after canonically normalising the vector field, we find
\begin{equation}\label{gauge2}
f = \tau.
\end{equation}
The total four dimensional action we will consider then takes the form
\begin{equation}
S = \int d^4x \sqrt{-g}\left(\frac{M_p^2}{2}R-\partial_\mu \chi \partial^\mu \chi - V(\chi)-\frac{1}{4}fF_{\mu\nu}F^{\mu\nu}-\frac{1}{2}m^2A_\mu A^\mu\right),
\end{equation}
where we have used Eq.~(\ref{sugra}) for a single canonically normalised inflaton $\chi$, and the relevant terms for the D7-brane vector field from the DBI and WZ actions. Note that for closed string inflation we do not consider the D-brane scalars here.

In an FRW background, the equation of motion for $\chi(t)$ becomes
\begin{equation}\label{tauEOM}
\ddot\chi + 3H\dot\chi + V'- \frac{1}{2}f'\left(\frac{\dot{A}}{a}\right)^2 + m m'\left(\frac{A}{a}\right)^2 =0,
\end{equation}
where we see that as for the D3-brane case, the vector field can backreact into the dynamics of the inflaton. The equations of motion for the vector background $A_\mu(t)$ are once again given by Eqs.~(\ref{vectorEOM1}) and (\ref{vectorEOM2}), where now the gauge kinetic function $f$ and mass $m$ are given by Eqs.~(\ref{gauge2}) and (\ref{7mass}) respectively.

Having all the equations and relevant quantities at hand, we will now consider whether or not such a set-up allows for an embedding of the vector curvaton scenario. Here we will check this using simple calculations that can be done analytically. From Eq.~(\ref{7mass}) we see that for $\mathcal{V}\gg 1$ we find $m \propto \tau^{-1/4}$, and because $f = \tau$, we immediately obtain Eq.~(\ref{a4case}). Therefore, the D7-brane curvaton can generate scale invariant spectra for all of its components if Eq.~(\ref{vectorcons2}) holds, \emph{i.e.} if $f\equiv \tau \propto a(t)^{-4}$. We will now investigate whether or not this is possible.

Let us consider the case for which the backreaction of the vector field into the dynamics of the inflaton is negligible (see \cite{Dimopoulos:2010xq} for a discussion of non-negligible backreaction). For $f = \tau \propto a^{-4}$, from Eq.~(\ref{canonicaltau}) we then require that the canonically normalised field obeys $\chi \propto a(t)^{-3}$. For inflation to occur we require $V\approx const.$, and indeed the potential in Eq.~(\ref{kahlerpot}) is exponentially flat for large enough $\tau$. For $H\approx const.$ and using the approximation $V' \approx 0$, the solution to Eq.~(\ref{tauEOM}) without backreaction takes the form\footnote{More precisely the full solution, which can only be obtained numerically using the full form of the potential, will contain suppressed terms that come from the dependence of $V'$ on $\chi$. Here we are interested only in the dominant contributions.}
\begin{equation}\label{solution}
\chi = -\frac{e^{-3Ht}}{3H}c_1 +c_2,
\end{equation}

We would like $c_2 \approx 0$ in which case the dominant solution is $\chi \propto e^{-3Ht} = a^{-3}$ as required. This means that firstly the minimum of the potential should occur at a \emph{very} small value for $\chi$ (this depends on the geometry) and secondly, $\chi$ should not approach the second term in Eq.~(\ref{solution}) until it is very close to its minimum (this depends on initial conditions). We therefore require that initially,
\begin{equation}
c_1 \ = \ \dot\chi_0 \ \lesssim \ -3H_0\chi_0.
\end{equation}
The value of $\chi$ should decrease during inflation, so we see that $c_1$ is negative as it should be. If $\chi$ is the inflaton, then we further require that $1/2\dot\chi^2 \ll V$ such that the energy density is roughly constant. From the Friedmann equation we obtain
\begin{equation}
3H_0^2\left(1-\frac{3}{2}\chi_0^2\right) = V(\chi)
\end{equation}
so we require $\chi_0 \ll \sqrt{2/3}$, \emph{i.e.} the field should be subPlanckian to begin with. Then, for 10 efolds of evolution which span the cosmological scales, we have $N = \ln(a_f/a_i) = -1/3\ln(\chi_f/\chi_i)$, so $\chi_f \approx 10^{-14}\chi_i$, which is rather a large field range. We would further require that the other 50 efolds of inflation take place over a very small field range, depending upon just how small the stabilised value of $\chi$ can be. This seems feasible given the vast number of Swiss cheese geometries, but it would have to be verified concretely, checking that the over-all volume and the other volume moduli are stabilised properly in the process, with realistic values for the parameters. If $c_2$ is a small constant, this will add a degree of dependence upon scale to the curvature perturbation, as can be seen from Eqs.~(\ref{powerLR},\ref{powerLong}) considering that $\alpha = 0$.

\section{Conclusions}
We have explored the possibility of embedding the vector curvaton paradigm in Type IIB string theory as a promising mechanism to generate statistical anisotropy, in both the open and closed string sectors. We first considered a toy model in open string inflation, where the vector field on a D3-brane plays the role of the vector curvaton. With simple computations we then demonstrated the feasibility of constructing a concrete model in closed string inflation, where the vector field on a D7-brane plays the role of the curvaton.

\end{document}